\documentstyle[aps,prl,multicol,epsf]{revtex}
\def\gs{\gtrsim}
\def\ls{\lesssim}
\def\be{\begin{equation}}
\def\en{\end{equation}}                  
\newcommand{\bi}[1]{\mbox{\boldmath$#1$}}
\newcommand{\av}[1]{\langle{#1}\rangle}

\begin{document}

\draft
\bibliographystyle{prsty}
\title{Heterogeneous Diffusion in Highly Supercooled Liquids}
\author{R. Yamamoto\cite{ryoichi} and A. Onuki\cite{onuki}}
\address{Department of Physics, Kyoto University, Kyoto 606-8502}
\date{\today}
\maketitle

\begin{abstract}
The diffusivity of tagged particles is demonstrated 
to be very heterogeneous  on time scales comparable to or shorter than 
the $\alpha$ relaxation time $\tau_{\alpha}$ ($\cong$ the stress 
relaxation time)  
in a highly supercooled liquid via 3D molecular dynamics simulation. 
The particle motions in the relatively active regions 
dominantly contribute to the mean square displacement,  
giving rise to a diffusion constant systematically larger than 
the Einstein-Stokes value.
The van Hove self-correlation function $G_s(r,t)$ is shown 
to have a long  distance tail 
which can be  scaled in terms of 
$r/t^{1/2}$ for $t \ls 3\tau_{\alpha}$. 
Its presence  indicates   heterogeneous  
diffusion  in the active regions.   
However, the diffusion process eventually becomes homogeneous 
on time scales longer than the life time of the heterogeneity 
structure ($\sim 3 \tau_{\alpha}$).
\end{abstract}

\begin{multicols}{2}

In a wide range of liquid states 
the Einstein-Stokes relation 
$D\eta a/ k_BT=const.$  has been  successfully 
applied  between the translational 
diffusion constant $D$ of a tagged particle 
and the viscosity $\eta$  even 
when the tagged particle diameter $a$ 
is of the same order as that of solvent molecules. 
However, this relation is systematically violated 
in  fragile  supercooled liquids \cite{Ediger,Sillescu,Ci95}.
The diffusion process  in  supercooled liquids thus  remains 
not well understood. 
In particular, Sillescu {\it et al.} 
observed the power law behavior 
$D \propto \eta^{-\nu}$ with $\nu \cong 0.75$ 
at low temperatures \cite{Sillescu}. 
Furthermore, Ediger {\it et al.} found that 
smaller probe particles  exhibit a more pronounced increase of  
$D\eta/T \propto D/D_{SE}$ 
with lowering $T$  \cite{Ci95}, 
where $D_{SE} \sim  k_BT/2\pi \eta a$ is the Einstein-Stokes  
diffusion constant. In such experiments 
the viscosity changes over 12 decades with 
lowering $T$, while the ratio  
 $D/D_{SE}$ increases from of order 1 up to order $10^2 \sim 10^3$.  
In molecular dynamics  simulations, on the other hand, 
  the same tendency has been   
detected  in a three dimensional (3D) binary mixture with $N=500$ particles 
\cite{Mountain}   and in a two dimensional (2D)
 binary mixture with $N=1024$ 
\cite{Perera_PRL98}. In our  recent 
3D simulation with $N= 10^4$ \cite{Yamamoto_Onuki98}, 
$\eta$ and $D$ have both varied    over 4 decades  and 
the power law behavior $D \propto \eta^{-0.75}$ has been observed. 
Many authors have attributed the origin of 
the breakdown to   heterogeneous coexistence of 
 relatively active  and inactive regions, 
 among which  the local  diffusion constant is expected 
 to vary significantly \cite{Sillescu,Ci95,St94,Tarjus,Oppen}.
The aim of this paper is to numerically 
demonstrate that the diffusivity 
of the particles is indeed very heterogeneous 
on time scales  shorter than the structural or $\alpha$ 
relaxation time but becomes homogeneous 
on  time scales much longer than $\tau_{\alpha}$.

A number of recent MD simulations have 
detected dynamic heterogeneities in supercooled model 
binary mixtures \cite{Muranaka,Harrowell,Yamamoto_Onuki1,Donati} 
to confirm 
a picture of  {\it cooperatively rearranging regions}  
 \cite{Adam}. That is, rearrangements of particle configurations in 
 glassy materials  are  cooperative, involving many molecules,  
 owing to configuration  restrictions.  
 In particular, we have examined  
 bond breakage processes among adjacent particle pairs 
and  found that the broken bonds in an  appropriate 
time interval ($\sim \tau_{\alpha}$) 
are very analogous to the critical fluctuations 
in Ising spin systems with their  structure factor being 
excellently fitted to the Ornstein-Zernike form 
\cite{Yamamoto_Onuki98,Yamamoto_Onuki1}.  
The correlation length $\xi$ thus obtained 
increases up to the system size and satisfies 
the dynamic scaling law,  $\tau_{\alpha} \sim \xi^z$,  with 
$z=4$ in 2D and $z=2$ in 3D. 
The heterogeneity structure in the bond breakage is essentially 
the same as that in jump motions of particles from cages 
or that in the local diffusivity, as will be discussed below.

Much attention has recently 
been paid to the mode coupling theory 
\cite{Gotze}.  
It   is a self-consistent 
scheme for the density time 
correlation function and describes  
onset of glassy slowing down or slow structural relaxations 
considerably above $T_g$.
However,  the mode coupling theory  predicts no long range 
correlations.

Our 3D  binary mixture is composed of two 
atomic species, $1$ and $2$, with $N_{1}=N_{2}=5000$ particles
with the system linear dimension  $L=V^{1/3}$ being fixed at 
$23.2\sigma_1$. They interact via the soft-core potentials  
$ v_{ab}(r)= \epsilon (\sigma_{ab}/r)^{12}$ with 
$\sigma_{ab}=(\sigma_{a}+\sigma_{b})/2$,   
where $r$ is the distance between two particles and 
$a,b= 1,2$ \cite{Bernu}.  The interaction is truncated at $r =3\sigma_{1}$.
The  mass ratio is  $m_{2}/m_{1}=2$. 
The size ratio is  $\sigma_{2}/\sigma_{1}=1.2$, 
 which prevents crystallization at least in our computation times.  
We fix  the particle density at a very high value of 
 $(N_1+N_2)/V =0.8/\sigma_{1}^{3}$,
so the particle configurations are severely jammed. 
We will measure space and time in units of 
$\sigma_1$ and  $\tau_0=({m_{1}\sigma_{1}^{2}/\epsilon})^{1/2}$. 
The temperature $T$ will be measured  in units of $\epsilon/k_B$,  
and the viscosity $\eta$ in units of $\epsilon \tau_0/\sigma_1^3$. 
Very long annealing times ($\sim 2.5\times10^5$) are  chosen  
in our case. Then,  for $T\ge0.267$  
 no appreciable aging effect is detected in various 
quantities such as the pressure or the density time 
correlation function, whereas  at the lowest temperature,  
$T=0.234$, a small aging effect remains  in the density 
time correlation function.

Let us consider  the incoherent  density correlation 
\begin{figure}[t]
\epsfxsize=3.0in
\centerline{\epsfbox{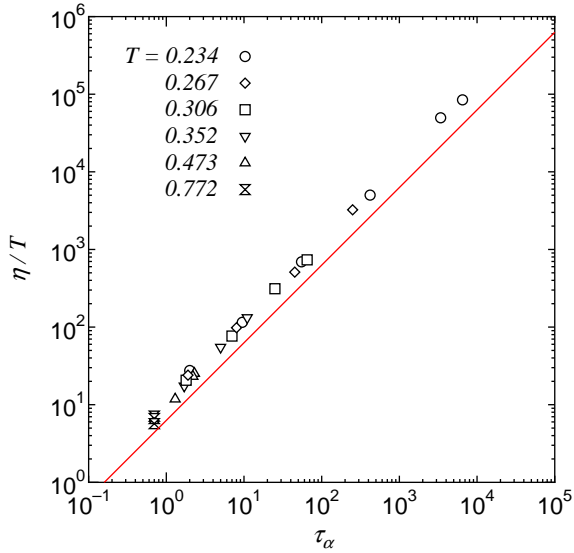}}
\caption{\protect\narrowtext
$\eta/T$ versus  $\tau_\alpha$ at various 
temperatures. 
Those are obtained from nonequilibrium MD 
in shear flow \protect\cite{Yamamoto_Onuki98}.
The straight line represents 
$\eta/T=2\pi \tau_\alpha$.}
\label{taua_vis}
\end{figure}
\begin{figure}[t]
\epsfxsize=3.0in
\centerline{\epsfbox{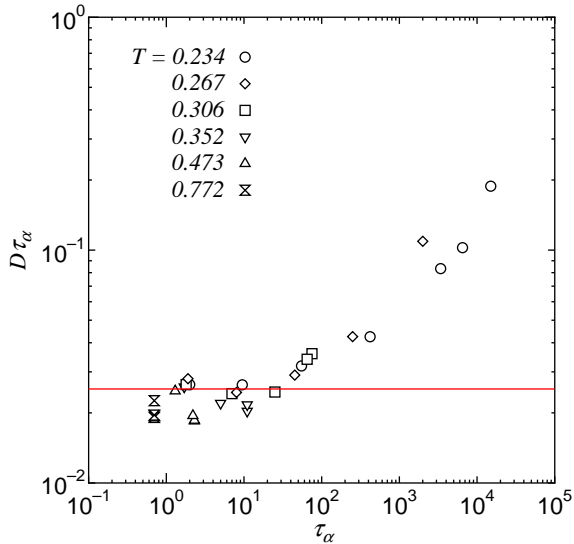}}
\caption{\protect\narrowtext
$D\tau_\alpha$ versus $\tau_\alpha$.
The solid line represents the Stokes-Einstein value 
$D_{ES}\tau_{\alpha}=(2\pi)^{-2}$ arising from Eq.(1).}
\label{taua_dtaua}
\end{figure}
\noindent
function, $F_s(q,t)= \av{\sum_{j=1}^{N_1}
\exp {[i{\bf q}\cdot \Delta{\bi r}_j(t)]}
}/N_1$ 
for the particle species 1, where 
$\Delta{\bi r}_j(t)={\bi r}_j(t)- {\bi r}_j(0)$ 
is the displacement vector of the $j$-th particle. 
This function may be introduced also in shear flow \cite{Yamamoto_Onuki98}. 
The $\alpha$ relaxation 
time $\tau_{\alpha}$  is then defined by   
$F_s(q,\tau_{\alpha})= e^{-1}$ at $q = 2\pi$ for various 
 $T$ (and the shear rate $\dot{\gamma}$). 
We also  calculate    the coherent time 
correlation function,  $S_{11}(q,t)= 
\langle 
n_1({\bi q}, t)n_1({-{\bi q}}, 0)
\rangle$,  
for the density fluctuations  of the  particle species 1.  
Interestingly, the decay profiles of   
$S_{11}(q,t)$ at its first peak 
wave number 
$q=q_m \sim 2\pi$ 
and $F_s(q,t)$ at $q=2\pi$ 
nearly coincide 
\begin{figure}[t]
\epsfxsize=3.0in
\centerline{\epsfbox{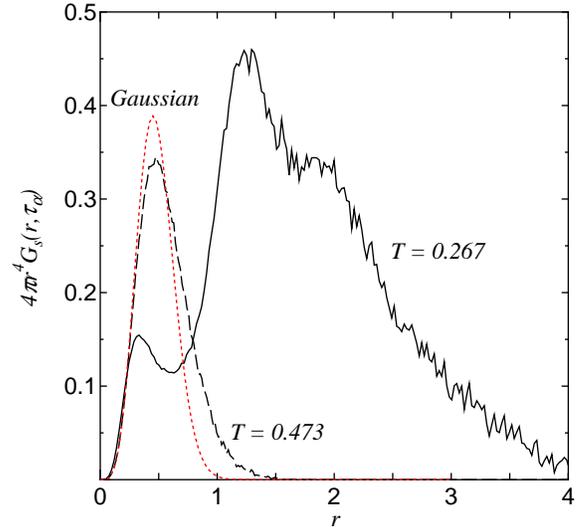}}
\caption{\protect\narrowtext
$4\pi r^4 G_s(r,t)$ versus $r$ at $t=\tau_{\alpha}$.
The solid line is for $T=0.267$ and the broken line is for $T=0.473$. 
The dotted line represents the  Brownian motion result.
Note that the areas below the curves give $6D\tau_{\alpha}$.
}
\label{6dta}
\end{figure}
\noindent
in the whole time region studied ($t < 2 \times 10^5$) 
within $5\%$. Hence $S_{11}(q_m,\tau_{\alpha})/S_{11}(q_m,0) \cong e^{-1}$ 
holds  for any $T$ in our simulation. 
Such agreement is not obtained for other wave numbers, however.
 Furthermore, some  neutron-spin-echo 
experiments \cite{Mezei} showed that 
 the decay time of $S_{11}(q_m,t)$ 
 is  nearly equal to  the stress 
relaxation time 
 and as a result the viscosity $\eta$ is of order $\tau_{\alpha}$.  
In accord with this experimental result, 
we obtain a simple linear relation in our simulation, 
\be
\tau_{\alpha}\cong  (2\pi\sigma_1/q_m^2)\eta /k_BT 
\label{eq:1}
\en
in the original units. 
 Fig.1  shows  that  Eq.(1) is valid for any 
$T$ and $\dot{\gamma}$ over a wide range of 
$\tau_{\alpha}$. 
Here we may define a $q$-dependent 
relaxation time  $\tau_q$ by $F(q,\tau_q)=e^{-1}$. 
Thus, at the peak 
wave number  $q=q_m$, 
 the effective diffusion constant   $D_q \equiv 1/q^2\tau_q$ 
is  given by  the Einstein-Stokes form  
  even in highly supercooled liquids.

However, notice that the usual diffusion constant  
is the long wavelength limit,  
$D= \lim_{q\rightarrow 0} D_q$. 
It is usually  calculated from the mean square  
displacement, 
$
\av{(\Delta{\bi r}(t))^2}= \av{
\sum_{j=1}^{N_1}(\Delta{\bi r}_j(t))^2}/N_1.
$
The crossover of this quantity  
 from the plateau behavior arising from  
 motions in transient cages to the diffusion behavior 
  $6Dt$ has been  found to take place around  $t \sim  0.1 \tau_{\alpha}$ 
  \cite{Yamamoto_Onuki98}. 
In Fig.2  we plot $D\tau_{\alpha}$ versus $\tau_{\alpha}$, 
which clearly indicates  breakdown of the Einstein-Stokes 
relation  in agreement with 
the experimental trend.  To examine the diffusion process 
in more detail we here  introduce   
the van Hove self-correlation function,  
$
G_{s}(r,t)=   
\av{ \sum_{j=1}^{N_1}\delta (\Delta{\bi r}_j(t)-{\bi r}) }/N_1 .
$  
Then,  
\be
F_s(q,t) = 
\int_0^\infty dr \frac{\sin(qr)}{qr}4\pi r^2 G_s(r,t)
\label{eq:2}
\en
is the 3D Fourier transformation of $G_{s}(r,t)$. 
At $q=2\pi$,  
\end{multicols}
\newpage
\widetext
\begin{figure}[t]
\centerline{
\epsfxsize=2.15in\epsfbox{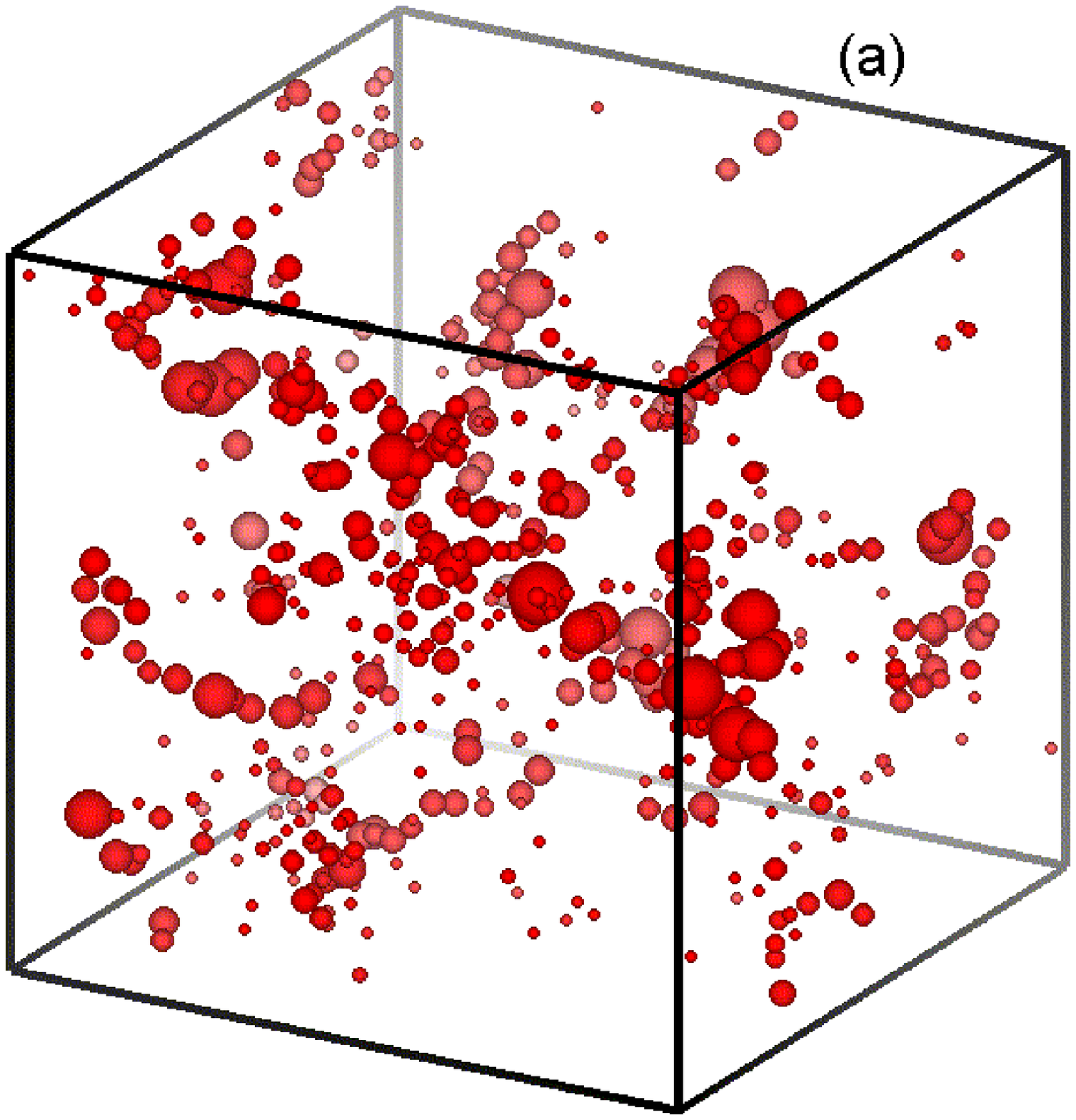}\hspace{5.5mm}
\epsfxsize=2.15in\epsfbox{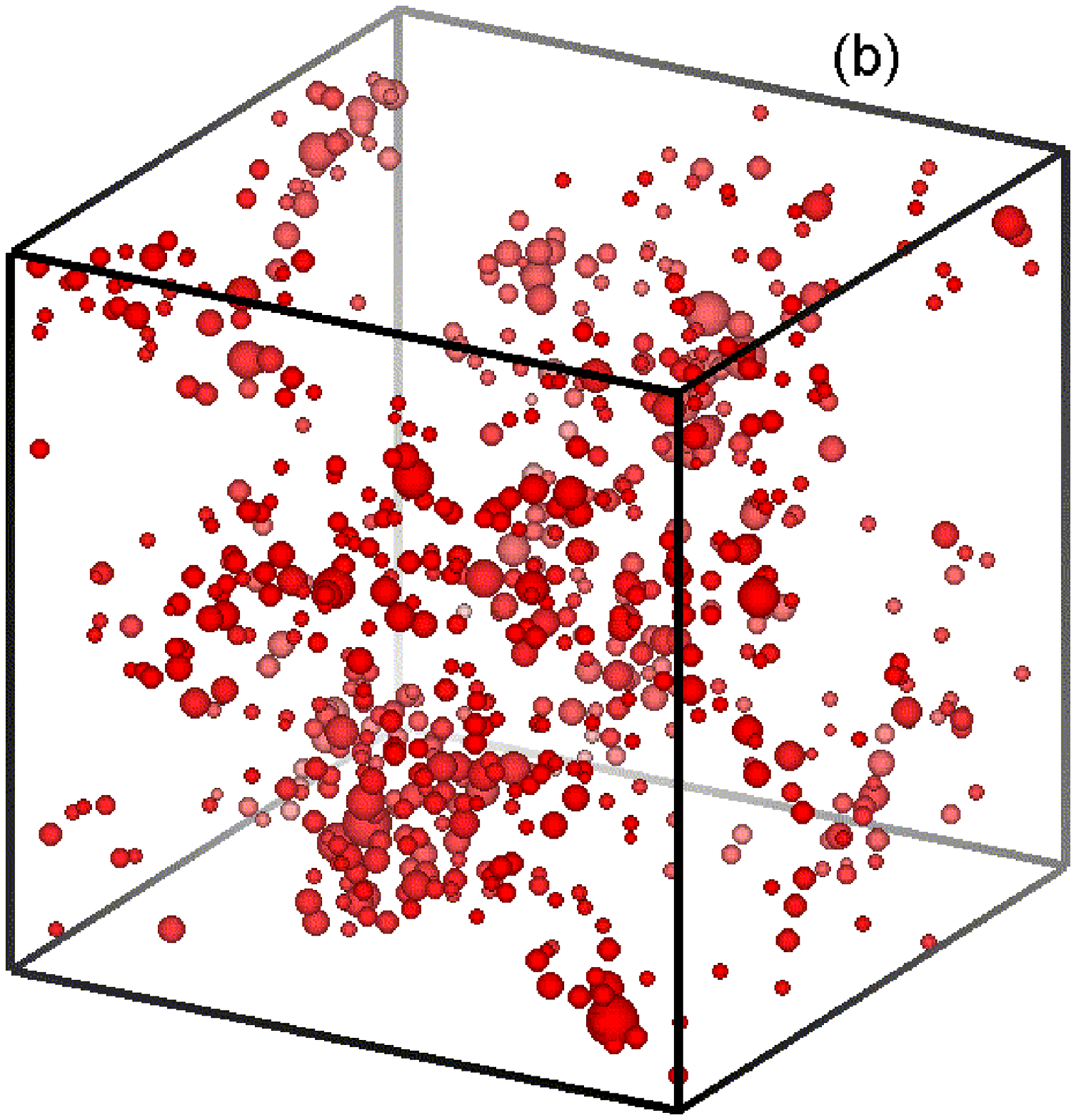}\hspace{5.5mm}
\epsfxsize=2.15in\epsfbox{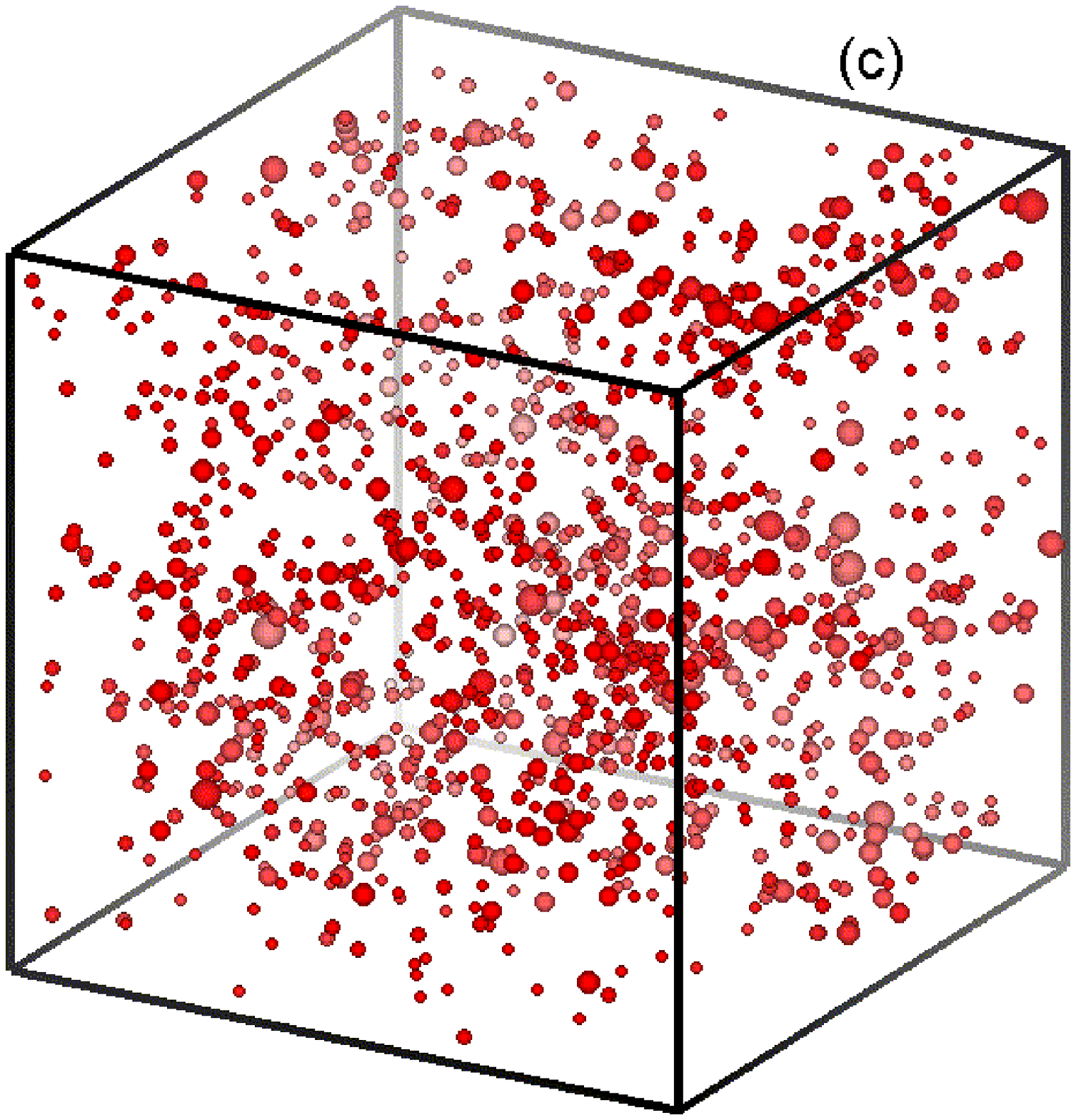}
}
\caption{
Mobile particles of the species 1 at $T=0.267$. 
The time interval $t$ is $0.125 \tau_{\alpha}$ in (a), 
$\tau_{\alpha}$ in (b), and  $10 \tau_{\alpha}$ in (c).  
The radii of the spheres are 
$|\Delta{\bi r}_j(t)|/\sqrt{\av{(\Delta{\bi r}(t))^2}}$ 
and the centers are at 
$\frac{1}{2} [ {\bi r}_j(t_0)+{\bi r}_j(t_0+t)]$.  
The system linear dimension is $L=23.2$.
The darkness of the spheres represents the depth in the 3D space.   
The heterogeneity is significant at $0.125 \tau_{\alpha}$ and $\tau_{\alpha}$ 
but is  much decreased at $10 \tau_{\alpha}$. 
}
\label{space}
\end{figure}
\begin{multicols}{2}
\noindent
the integrand in Eq.(2)  vanishes at $r=1$  
and the integral in the region $r<1$ is confirmed to 
dominantly determine 
the decay of $F_s(2\pi,t)$ or $\tau_{\alpha}$. On the other hand, the 
mean square displacement 
\be
\av{(\Delta{\bi r}(t))^2}= 
\int_0^\infty dr 4\pi r^4 G_s(r,t)
\label{eq:3}
\en  
is determined by the particle motions out of  the cages. 
In Fig.3 we display $4\pi r^4 G_s(r,\tau_{\alpha})$ versus 
$r$ at zero shear, 
where $\tau_{\alpha}=3.2$ and $2000$   for $T=0.473$ and $0.267$, 
respectively. 
These curves may be compared with 
 the Gaussian (Brownian motion)   result, 
$(2/\pi)^{1/2}\ell^{-3}r^4 \exp (-r^2/2\ell^2)$,  
where $3 \ell^2= 6D_{ES}\tau_{\alpha} = 3/2\pi^{2}$ is the 
Einstein-Stokes mean square displacement. 
Because the areas below the curves 
give $6D\tau_{\alpha}$, we recognize  that 
the particle motions over large distances $r > 1$  
are much enhanced  at low $T$, leading  to the violation 
of the Einstein-Stokes relation.

We then visualize the heterogeneity of the diffusivity.
To this end,  we pick up  mobile particles of the species 1 
with   $|\Delta{\bi r}_j(t))| > \ell_c$ 
in a time interval $[t_0, t_0+t]$ and number them as 
$j=1, \cdots, N_m$. 
Here $\ell_c$ is defined such that the sum of 
$\Delta{\bi r}_j(t)^2$ of the mobile particles is $66\%$ of the total sum 
($\cong 6DtN_1$ for $t \gs 0.1\tau_{\alpha}$).
In Fig.4 these particles are written  as  spheres 
with radius 
\be
a_j(t) \equiv |\Delta{\bi r}_j(t)|/\sqrt{\av{(\Delta{\bi r}(t))^2}}
\label{eq:4}
\en 
located at ${\bi R}_j(t)\equiv  
\frac{1}{2} [ {\bi r}_j(t_0)+{\bi r}_j(t_0+t)]$
in three time intervals, 
$[t_0, t_0+0.125\tau_{\alpha}]$ in (a), $[t_0, t_0+\tau_{\alpha}]$ in (b), 
and $[t_0, t_0+10\tau_{\alpha}]$ in (c).  The lower cut-off 
$\ell_c$ and  the mobile 
particle number $N_m$ increase with increasing $t$ as 
$0.48$ and $571$ in (a), $1.3$ and $725$ in (b), and 
$2.9$ and $1316$ in (c), respectively.  
Here they approach the Gaussian  results, 
 $\ell_c = 0.403 (t/\tau_{\alpha})^{1/2}$ 
and $N_m=1800$,  for $t \gg \tau_{\alpha}$. 
In our case  the ratio of the second moments 
$c_2 \equiv \sum_{j=1}^{N_m} a_j(t)^2/\sum_{j=1}^{N_1} a_j(t)^2$ 
is held fixed at $0.66$ independently of $t$, while  
the ratio of fourth moments 
$c_4 \equiv \sum_{j=1}^{N_m} a_j(t)^4/\sum_{j=1}^{N_1} a_j(t)^4 $ 
turns out to be close to 1 as  $c_4=0.97$ in (a), $0.92$ 
in (b), and $0.90$ in (c). 
We can see that the large scale heterogeneities in (a) and (b) 
are much weakened in (c) and that 
 the  areas of the spheres have the largest  variance in (a).  In fact, 
 the variance defined by ${\cal V} \equiv 
 N_m \sum_{j=1}^{N_m} a_j(t)^4/(\sum_{j=1}^{N_m} a_j(t)^2)^2  -1 $
is $0.94$ in (a), $0.41$ in (b), and $0.32$ in (c). 
Here  $c_4 \rightarrow 0.833$ and ${\cal V} \rightarrow 0.13$ 
  for $t \gg \tau_{\alpha}$.
The above result is consistent  with the fact that the 
 non-Gaussian parameter $A_2(t)$ 
 takes a maximum of 3.1 at $t \cong 0.125 \tau_{\alpha}$ at this temperature.
This is because 
  the statistical average  of 
$\cal V$ (taken over many initial times $t_0$) is related to  $A_2(t)$  
by $\av{{\cal V}} \cong (5 \av{c_4}\av{N_m}/3c_2^2 N_1) (1+ A_2(t)) - 1$. 
We may also conclude that the  significant rise of 
$A_2(t)$ in glassy states originates from 
the heterogeneity in accord with some 
 experimental interpretations  \cite{Zorn}.

Furthermore, 
we consider 
the Fourier component of 
the {\it diffusivity} density defined by  
\be
{\cal D}_{{\bi q}}(t_0, t) \equiv 
\sum_{j=1}^{N_m} a_j(t)^2 \exp [i{\bi q}\cdot ({\bi r}-{\bi R}_j(t))],
\label{eq:5}
\en
which depends  on the initial time $t_0$ and the final time 
$t_0+t$. The correlation function 
 $S_{{\cal D}}(q,t, \tau) = 
 \av{{\cal D}_{{\bi q}}(t_0 +\tau, t){\cal D}_{-{\bi q}}(t_0 , t)}$ 
is then obtained after averaging over many initial states. 
We  confirm that $S_{\cal D}(q,t, \tau)$ tends to 
its  long wavelength  limit  
for $q \ls \xi^{-1}$, where $\xi$ coincides with 
the correlation length of the heterogeneity 
structure of the bond breakage 
\cite{Yamamoto_Onuki98,Yamamoto_Onuki1}.   
As the difference $\tau$ of the initial times 
increases with fixed $t=\tau_{\alpha}$,   
$S_{{\cal D}}(q, \tau_{\alpha}, \tau)$  relaxes as 
$\exp [ - (\tau / \tau_h )^c ]$   
for $q \ls \xi^{-1}$, 
where $c \sim 0.5$ at $T=0.267$ and 
$\tau_h \sim 3 \tau_{\alpha}$ is 
the life time of the heterogeneity  structure. 
The two-time correlation function
 among  the broken bond density 
  \cite{Yamamoto_Onuki1,Yamamoto_Onuki98} 
 also  relaxes  with $\tau_h$ in 
  the same manner.

We naturally expect that the distribution of the particle displacement 
$\Delta{\bi r}_j(t)$ in the active regions
should be characterized by the local 
diffusion constant $D({\bi x},t)$ dependent on the spatial 
position ${\bi x}=(x,y,z)$ and  the 
\begin{figure}[b]
\epsfxsize=3.0in
\centerline{\epsfbox{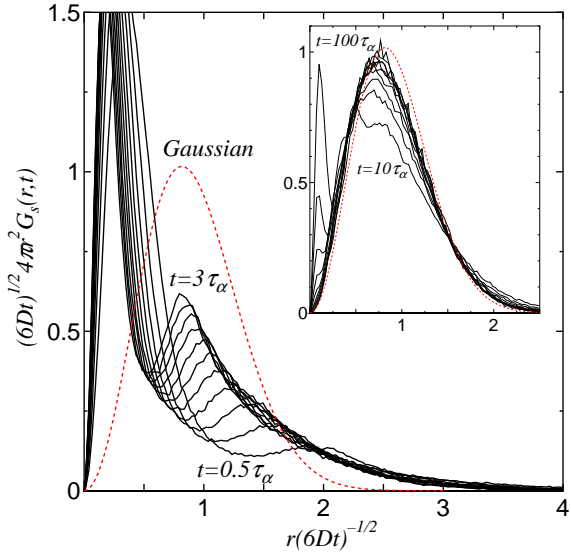}}
\caption{\protect\narrowtext
A test of the scaling plot 
$\sqrt{6Dt} 4\pi r^2 G_s(r,t)$ versus $r/\sqrt{6Dt}$ 
for $0.5 \tau_{\alpha} < t < 3\tau_{\alpha}$ ($t=(0.5+0.25n)\tau_{\alpha}$, $n=0,1,2,\cdots,10$).
The inset shows the curves at longer times, $10\tau_{\alpha} <t<100\tau_{\alpha}$ 
($t=10n\tau_{\alpha}$, $n=1,2,\cdots,10$), 
where the heterogeneity effect is smoothed out. 
The dotted lines are the Gaussian form. }
\label{scale}
\end{figure}
\noindent
time interval $t$. The van Hove correlation function 
$G_s(r,t)$ may then be expressed  as 
the spatial average of a local function $G_s({\bi x},r,t)$, which is 
given by 
$[4\pi D({\bi x},t)t]^{-3/2} 
\exp [ -{r^2}/4D({\bi x},t)t ]$ 
in the relatively  active regions. 
To check this conjecture we 
plot the scaled function $\sqrt{6Dt} 4\pi r^2 G_s(r,t)$ versus 
$r^*= r/\sqrt{6Dt}$ in Fig.5. The areas below the curves 
are fixed at 1. 
At relatively short times $t \ls 3 \tau_{\alpha}$, the curves 
in the region  $r \gs 1$ or $r^* \gs (6Dt)^{-1/2}$,
which give dominant contributions to $\av{(\Delta{\bi r}(t))^2}$,
tend to a master curve  
quite different from the rapidly decaying Gaussian tail.  
Note that the peak position of each curve at larger $r^*$ 
corresponds to $r \cong 1$ in Fig.5. 
This asymptotic law is consistent with the picture of the 
space-dependent diffusion constant in the active 
regions. It is also important that 
the heterogeneity structure 
remains unchanged in the time region $t \ls \tau_h \sim 3\tau_{\alpha}$.   
At longer times $t \gs 10\tau_{\alpha}$ the curves approach the 
Gaussian form as can be seen in the inset of Fig.5.
Of course,  
$4\pi r^2G_s(r,t)$ for $r<1$ 
does not  scale  in the above manner,  
because it is the probability density of  a tagged particle 
staying  within a cage.  This short-range behavior  
 determines the decay of $F_s(2\pi,t)$ as noted below Eq.(2).

We thank Professor T. Kanaya for helpful discussions. 
This work is supported by Grants in Aid for Scientific 
Research from the Ministry of Education, Science and Culture.
Calculations have been carried out 
at the Supercomputer Laboratory, 
Institute for Chemical Research, Kyoto University.

\end{multicols}
\end{document}